\documentclass{ragtime} 
\usepackage{longtable}
\usepackage{subfigure}
\title[Spectrum of $m=0$  horizontal disk oscillations]
{Frequency spectrum of axisymmetric horizontal oscillations in accretion
 disks}
\author[L. Giussani, 
        B. Mishra and
        W. Klu\'zniak]
       {L. Giussani\at{1} 
        W. Klu\'zniak\at{2}
        B. Mishra\at[]{2} \\
        \ins{1}Centre des Etudes Sup\'erieures Industrielle, 93 Boulevard de la Seine,\\ 92000 Nanterre, France\\
 \\
        \ins{2}Nicolaus Copernicus Astronomical Center, Bartycka 18,\\ 00-716 Warsaw, Poland }

\providecommand{\dif}{\mathrm{d}} 

\begin{document}

\begin{abstract}
We present the spectrum of eigenfrequencies of axisymmetric
acoustic-inertial oscillations of thin accretion disks for a
Schwarzschild black hole modeled with a pseudo-potential.  There are
nine discrete frequencies, corresponding to trapped modes.  Eigenmodes
with nine or more radial nodes in the inner disk belong to the continuum, whose
frequency range starts somewhat below the maximum value of the radial
epicyclic frequency.  The results are derived  under the assumption
that the oscillatory motion is parallel to the midplane of the disk. 
\end{abstract}

\section{Acoustic-inertial modes}
We consider acoustic-inertial modes of oscillation in the inner part
of an accretion disk, closely following the formalism of 
\citet{nowak1991,nowak1992}. Trapping of the fundamental axisymmetric
mode with no nodes in the vertical ($z$) direction was
first demonstrated by \citet{fukato1980}  in the Schwarzschild
geometry. \citet{nowak1991} derive the equations of motion in a
Lagrangian  pseudo-Newtonian formalism and specialize to purely
horizontal perturbed motions of the disk deriving eigenmodes 
and eigenfrequencies for the $m=0$ (axisymmetric)
and $m=2$ (quadrupole) modes.  \citet{shouryazofia} computed  in an improved pseudo-potential the
lowest radial modes (with up to three radial nodes) for azimuthal numbers $m=0$ through
$m=4$. Here, we present the complete
spectrum of horizontal axisymmetric acoustic-inertial disk modes  in
a pseudopotential which reproduces the properties of the
Schwarzschild-metric epicyclic frequency  \citep{kl2002,shouryazofia}.
The eigenfrequencies could be related to the quasi-coherent frequencies
 (QPOs) observed in the X-ray flux from black hole and neutron
star systems \citep[for a review see][]{vdk00},
as well as in cataclysmic variables
\citep[, and references therein]{woudt02}.

\section{Equation of motion and the boundary condition}
\begin{figure}[b]
\begin{center}
\includegraphics[width=.9\linewidth,height=.63\linewidth]{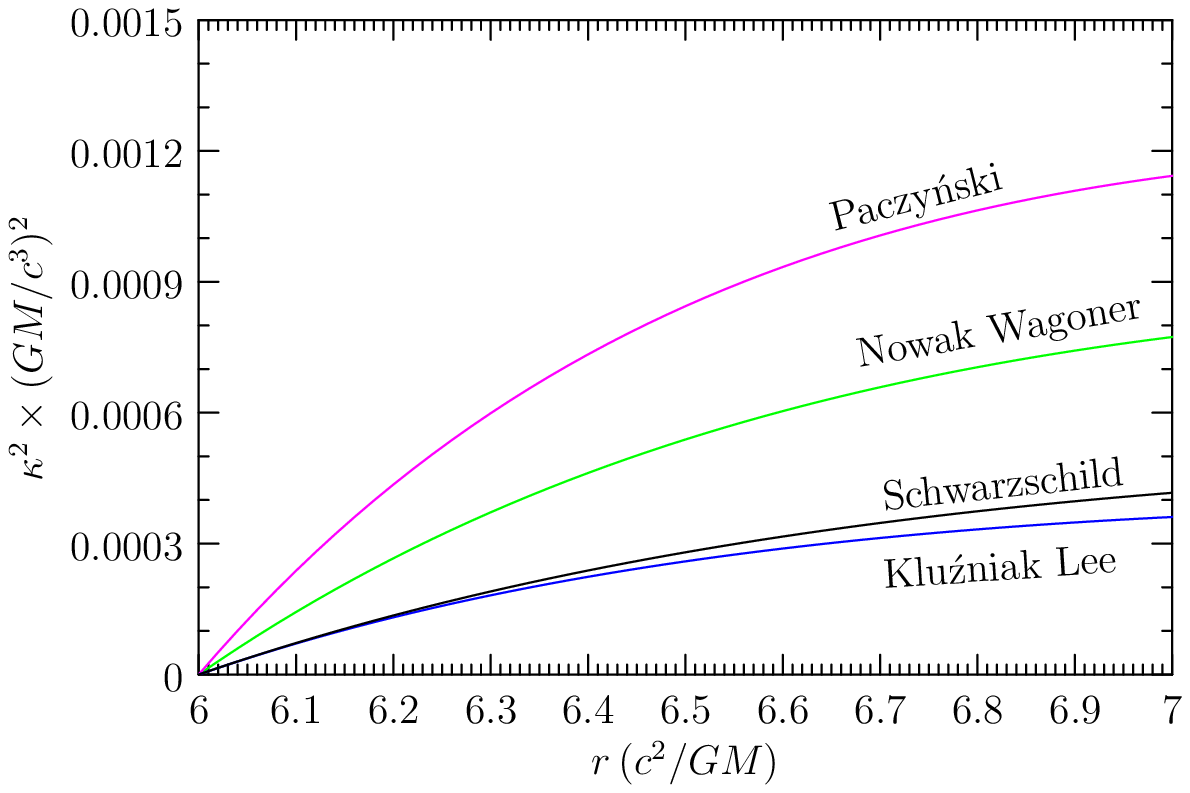}
\end{center}
\par\vspace{-1ex}\par
\caption{The Schwarzschild epicyclic frequency (squared)
 and its Newtonian models, from top to bottom:
\citet{PW1980,nowak1991}, our Eq.~(\ref{kl}).}
\label{kappasq}
\end{figure}

 We model the Schwarzschild metric
with a Newtonian pseudo-potential that reproduces the
Schwarzschild ratio of $\kappa^2(r)/\Omega^2(r)=1-6GM/(rc^2)$:
\begin{equation}
\Phi_{\rm KL}(r)=-(c^2/6)\exp\left(\frac{6GM}{rc^2}-1\right).
\label{kl}
\end{equation}
We have dropped an additive constant and renormalized the
original \citet{kl2002}  potential by a factor of $1/e$
to guarantee the correct Schwarzschild value of $\Omega(r_{\rm ms})$.
The orbital frequency can be obtained from
   $\Omega^2(r)=r^{-1}\partial\Phi_{\rm KL} /\partial r$, 
the radial epicyclic frequency from 
$\kappa^2=(2\Omega/r)\dif (r^2\Omega)/\dif r$
and the marginally stable orbit is at the zero of $\kappa$, at
$r_{\rm ms}=6GM/(rc^2)$.
Fig.~\ref{kappasq} compares our $\kappa^2(r)$ with the Schwarzschild
form and two other well-known pseudo-Newtonian models
\citep{PW1980,nowak1991}.

In this contribution we assume axisymmetric ($m=0$) horizontal
modes, with the perturbation vector in cylindrical coordinates
$(\xi_{*}^r,\xi_{*}^{\phi},\xi_{*}^z)
=(\xi^r,\xi^{\phi},0)\exp(i\sigma t)$.
We use the  equation of motion for
$\Psi(r)\equiv \sqrt{\gamma P r}\,\xi^r(r)$ derived
  in the Lagrangian formalism of  \citet{friedman78} by
\citet{nowak1991},
$$
{c_{\rm s} ^{2}}{\dif ^{2} \Psi}/{\dif r^{2}} +
{(\sigma^{2} - \kappa ^{2})}\Psi = 0,
$$
%
who also show that in the WKB approximation
the azimuthal component of the equation of perturbed
motion for thin disks reduces to
$\xi ^{\phi} = 2i \big({\Omega}/{\sigma}) \xi ^{r}$.

Following \citet{shouryazofia} we rewrite the equation of motion, and
the boundary condition that the Lagrangian perturbation of pressure
vanishes at the unperturbed boundary, in dimensionless form
as
\begin{equation}
\frac{\dif ^{2} \Psi}{\dif x^{2}} +
  \left(\frac{a}{H} \right)^2\left(\tilde\sigma ^{2}
 - \tilde\kappa ^{2}\right)\Psi = 0,
\label{prime}
\end{equation}
with the boundary condition at $x=0$
\begin{equation}
\frac{\dif \Psi}{\dif x} =
 - \frac{\Psi}{2},
\label{bc}
\end{equation}
where $H$ is the half-thickness of the disk, $a=r_{\rm ms}$,
and the dimensionless variables are given by
$r=a(1+x)$,
$\tilde\sigma =\sigma/\Omega(a)$, 
$\tilde\kappa(x) =\kappa(r)/\Omega(a)$.
The speed of sound $c_{\rm s}=\sqrt{\gamma P/\rho}$ was eliminated
with the condition of vertical hydrostatic equilibrium.
\begin{figure}[b]
\begin{center}
\includegraphics[width=.45\linewidth,height=.45\linewidth]{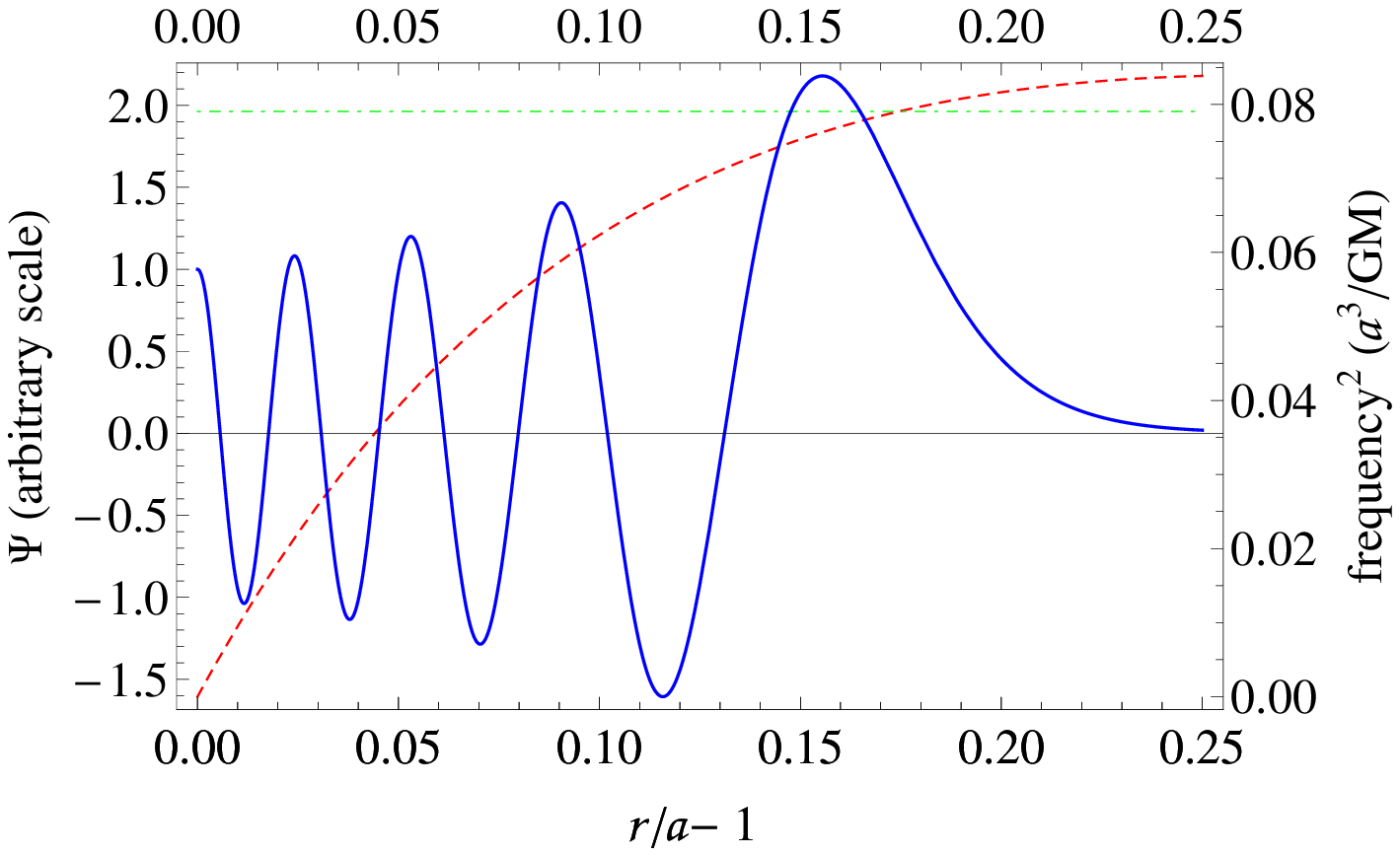}
\includegraphics[width=.45\linewidth,height=.45\linewidth]{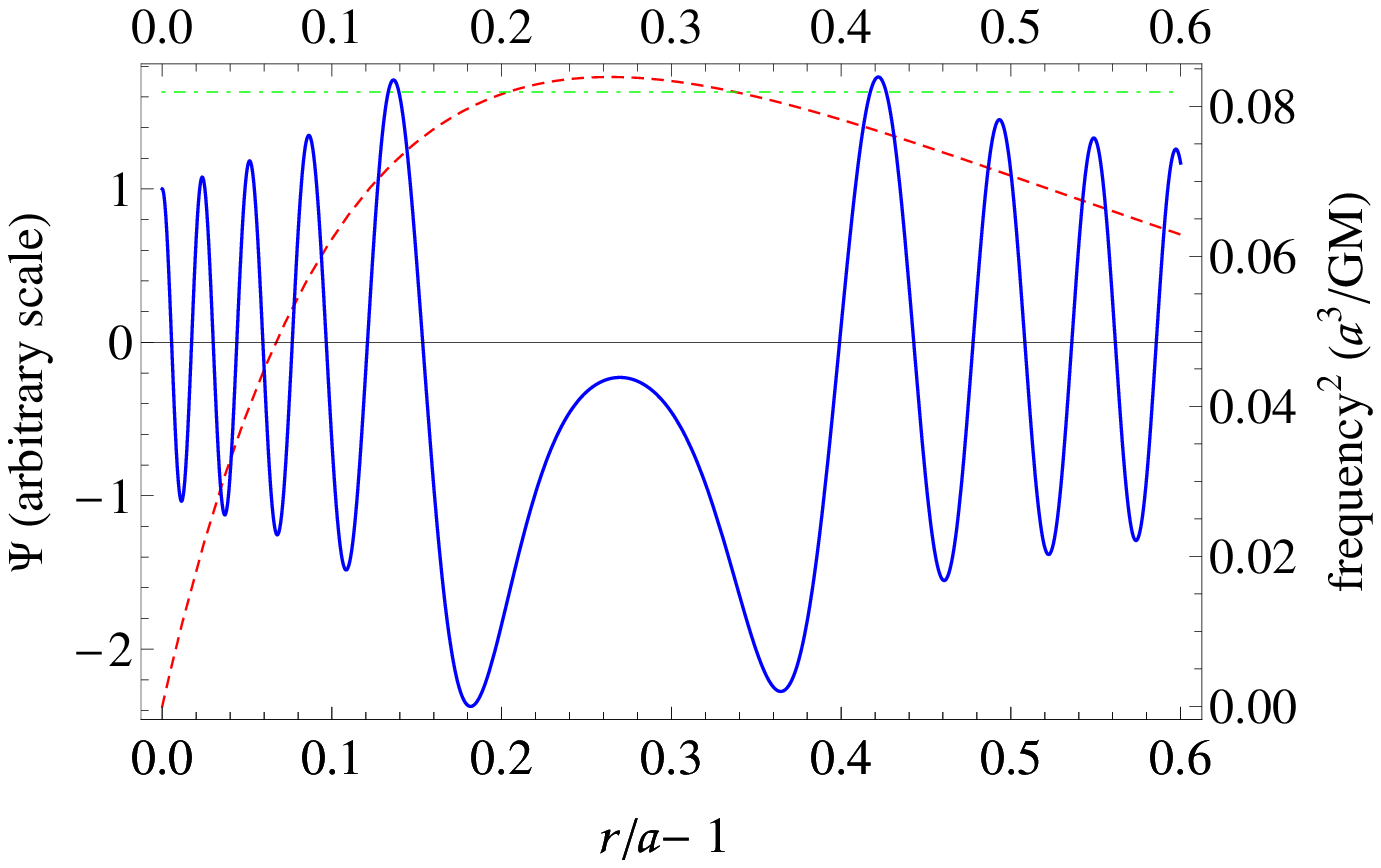}
\end{center}
\par\vspace{-1ex}\par
\caption{Two radial overtones for $m=0$
 horizontal oscillations of a thin
($H/a=10^{-3}$) accretion disk.
{\sl Left Panel:} A trapped oscillation with $\mu=8$ radial nodes.
{\sl Right Panel:} An oscillation penetrating the epicyclic barrier
(with $\mu=9$ radial nodes in the inner accretion disk and an unlimited number
of radial nodes in the outer disk).
Plotted are the  radial wavefunction $\Psi\propto r^{1/2}\xi^r$:
solid  blue line (left scale);
 eigenfrequency (squared) and the epicyclic frequency (squared),
both normalized to orbital frequency at the inner edge of the disk,
$\sigma^2/\Omega^2(r_{\rm ms})$: dashed-dotted green line,
$\kappa^2(r)/\Omega^2(r_{\rm ms})$:
dashed red line (right scale).}
\label{waves}
\end{figure}

\section{The eigenfrequency spectrum}

\begin{figure}[b]
\begin{center}
\includegraphics[width=\linewidth,height=.7\linewidth]{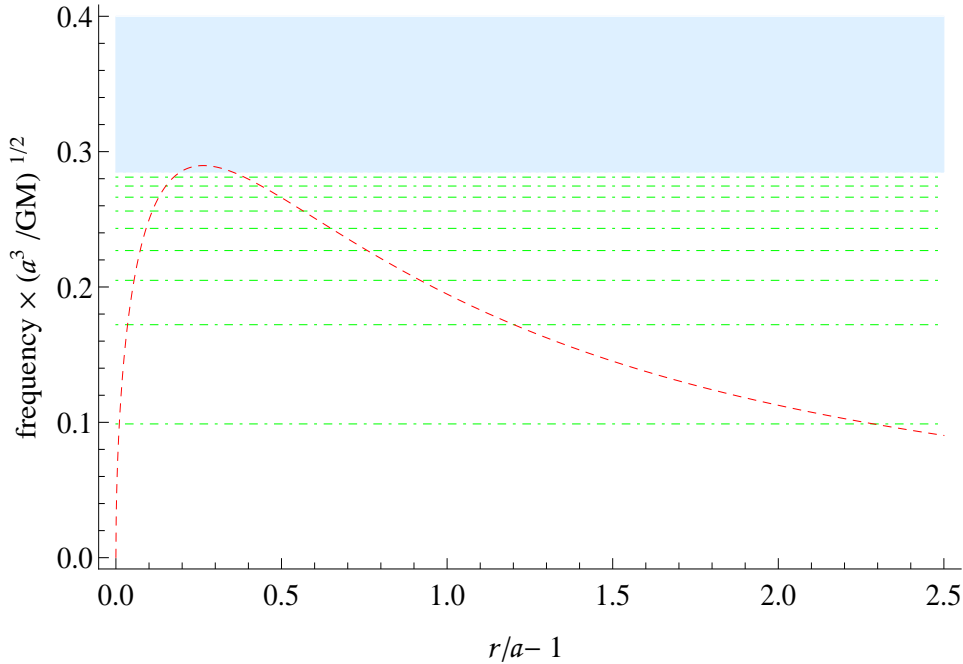}
\end{center}
\par\vspace{-1ex}\par
\caption{The spectrum of $m=0$ (axially symmetric)
 horizontal oscillations of a thin
($H/a=10^{-3}$) accretion disk for the potential of Eq.~(\ref{kl}).
 Plotted are the  epicyclic frequency (dashed red curve),
and the eigenfrequencies
$\sigma$ in the discrete set (dashed-dotted green lines)
and in the continuum (shaded blue region).
All frequencies were scaled with $(GM/a^3)^{1/2}$. 
Here, and throughout the paper, $a=r_{\rm ms}$.}
\label{spectrum}
\end{figure}

We have numerically solved the eigenvalue problem given by
 Eqs.~(\ref{prime}), (\ref{bc}), for a thin disk of $H/a=10^{-3}$,
and present  in Table~1 the eigenfrequencies for modes
with $\mu=0, 1,...,9$ radial nodes in the inner disk.
 The lowest nine eigenfrequencies
($\mu=0$ through 8), exhausting the discrete spectrum,
 correspond to oscillations which are trapped
in the inner disk. As already noted by \citet{fukato1980},
for eigenfrequencies exceeding the maximum of the epicyclic frequency,
$\sigma^2>\kappa^2_{\rm max}$, the acoustic wave ranges throughout the disk
\citep[see also][]{bluebook},
these frequencies belong to the continuum spectrum.
The tenth entry in Table~1, with $\mu=9$ radial nodes
in the inner disk, also belongs to the continuum, although it has
a frequency below the maximum of the epicyclic frequency
$\sigma^2<\kappa^2_{\rm max}$ (Figs.~\ref{waves},~\ref{spectrum}). 

 In Fig.~\ref{waves} we present the two eigenmodes corresponding to
the last two entries in Table~1.
The equations being linear in $\Psi$,
we normalize the wavefunction to unity at the inner edge of the disk:
 $\Psi(r_{\rm ms})=1$ for illustration purposes.
The left panel shows the highest-frequency eigenmode
in the discrete portion of the spectrum of axisymmetric ($m=0$)
horizontal disk oscillations, the wavefunction of this mode has
$\mu=8$ radial nodes.
 Note that the wave becomes evanescent for 
$\sigma^2<\kappa^2$, thus trapping the $\mu=8$ mode to the left of 
$\kappa_{\rm max}$. Some of the lower overtones have been illustrated
in  \citet{shouryazofia}.

The right panel of Fig.~\ref{waves} illustrates one of the lowest frequency
modes in the continuum. Here, $\sigma<\kappa_{\rm max}$ and is so close
in value to $\kappa_{\rm max}$ that the wave is transmitted through the 
epicyclic barrier to the outer disk, where it has an unlimited number of
radial nodes in addition to the  $\mu=9$ radial nodes
in the inner disk. As far as we are aware, this is a new finding,
which has never been reported before.
It may have an interesting astrophysical consequence. If the oscillations
arise close to the marginally stable orbit, as suggested by
 \citet{1987Natur.327..303P},
the ones transmitted to the outer disk are likely to be more easily
observable, in that they may modulate the emission from large parts
of the disk.

\begin{center}
\begin{longtable}{c|c|c|c|c}
\caption{Spectrum of eigenfrequencies for horizontal modes}\\
\hline
mode & radial nodes   & eigenfrequency & mode &\\
 $m$ & $\mu$ & $\sigma\times\sqrt{(a^{3}/GM)}$ & status & ref.\\
\hline	
0 & 0 &  0.098829oooo& trapped & \citet{shouryazofia}\\
0 & 1 &  0.172137256o& trapped & \citet{shouryazofia} \\
0 & 2 &  0.2049091375& trapped & \citet{shouryazofia}\\
0 & 3 &  0.22691222oo& trapped & \citet{shouryazofia}\\
0 & 4 &  0.2433069ooo& trapped & this work \\
0 & 5 &  0.256113oooo& trapped & this work \\
0 & 6 &  0.266346oooo& trapped & this work \\
0 & 7 &  0.274583oooo& trapped & this work \\
0 & 8 &  0.2811589ooo& trapped & this work \\
0 & 9 &  0.2862229ooo& not trapped & this work \\
\hline
\end{longtable}
\label{eigekl}
\end{center}

\section{Discussion}
We consider accretion disk oscillations in
a Newtonian model of the Schwarzschild metric, Eq.~(\ref{kl}),
 which accurately
models the radial epicyclic frequency, at least close to the marginally
stable orbit, see Fig.~\ref{kappasq}. No model is perfect, so although
we correctly reproduce the ratio of epicyclic to orbital
frequency $\kappa(r)/\Omega(r)=\sqrt{1-6GM/(rc^2)}$, and the correct value
of orbital frequency at the marginally stable orbit,
 $\Omega(r_{\rm ms})=c^3/(\sqrt{216}\,GM)$, the maximum of $\kappa$
occurs at $r=(3+\sqrt{21})(GM/c^2)\approx 7.58(GM/c^2)$ instead of
the Schwarzschild value $r=8(GM/c^2)$. Further, the equations of motion
for the oscillation of the disk fluid
were derived in a Newtonian formalism, not in full GR. These
departures from GR may limit the quantitative accuracy of
the presented results when applied to real black hole (or neutron star)
accretion disks. An additional assumption which may not be quite accurate
is that the oscillations of the disk are strictly parallel to the
midplane of the disk, i.e., that the perturbation vector has a zero
vertical component, $\xi^z=0$.

We find that the  spectrum of horizontal oscillations is composed of
nine discrete frequencies and a continuum (Fig.~\ref{spectrum}). For
the discrete spectrum the wave propagation region corresponds to those
regions where $\sigma^2>\kappa^2(r)$ and is separated into the inner
region of trapped oscillations,
from $r=r_{\rm ms}$ to $r\approx (7/6)\,r_{\rm ms}$, and an outer
region extending to $r>>r_{\rm ms}$ \citep{bluebook}.  However, the
lowest frequency modes in the continuum, which satisfy
$\sigma<\kappa_{\rm max}$, are transmitted through the epicyclic
barrier, and thus fluctuations in the inner disk may be transmitted to
the outer disk for frequencies close to the maximum of the epicyclic
one, $\sigma^2\approx\kappa^2_{\rm max}$.


\ack
This work was supported in part
by Polish NCN grant 2013/08/A/ST9/00795.

\bibliography{pgius}

\begin{thebibliography}{11}
\expandafter\ifx\csname natexlab\endcsname\relax\def\natexlab#1{#1}\fi
\expandafter\ifx\csname url\endcsname\relax
  \def\url#1{\texttt{#1}}\fi
\expandafter\ifx\csname urlprefix\endcsname\relax\def\urlprefix{URL }\fi
\providecommand{\selectlanguage}[1]{\relax}
\providecommand{\eprint}[2][]{\url{#2}}

\bibitem[{Friedman and Schutz(1978)}]{friedman78}
Friedman, J.~L. and Schutz, B.~F. (1978), Lagrangian perturbation theory of
  nonrelativistic fluids, \emph{ApJ}, \textbf{221}, p. 937.

\bibitem[{Kato and Fukue(1980)}]{fukato1980}
Kato, S. and Fukue, J. (1980), Trapped radial oscillations of gaseous disks
  around a black hole, \emph{PASJ}, \textbf{32}, p. 377.

\bibitem[{Kato et~al.(1998)Kato, Fukue and Mineshige}]{bluebook}
Kato, S., Fukue, J. and Mineshige, S. (1998), {Black-Hole Accretion Disks},
  \emph{Kyoto University Press}.

\bibitem[{Khanna et~al.(2014)Khanna, Strzelecka, Mishra and
  Klu\'zniak}]{shouryazofia}
Khanna, S., Strzelecka, Z., Mishra, B. and Klu\'zniak, W. (2014), Eigenmodes of
  trapped horizontal oscillations in accretion disks, \emph{Proc. of RAGtime,
  eds. S. Hled\'ik and Z. Stuchl\'ik}, p. in press., arXiv:1411.4690

\bibitem[{Klu\'zniak and Lee(2002)}]{kl2002}
Klu\'zniak, W. and Lee, W.~H. (2002), The swallowing of a quark star by a black
  hole, \emph{MNRAS}, \textbf{335}, p. L29.

\bibitem[{Nowak and Wagoner(1991)}]{nowak1991}
Nowak, M.~A. and Wagoner, R.~V. (1991), {Diskoseismology: Probing accretion
  disks. I - Trapped adiabatic oscillations}, \emph{ApJ}, \textbf{378}, p. 656.

\bibitem[{Nowak and Wagoner(1992)}]{nowak1992}
Nowak, M.~A. and Wagoner, R.~V. (1992), {Diskoseismology: Probing accretion
  disks. II - G-modes, gravitational radiation reaction, and viscosity},
  \emph{ApJ}, \textbf{393}, p. 697.

\bibitem[{{Paczy\'nski}(1987)}]{1987Natur.327..303P}
{Paczy\'nski}, B. (1987), {Possible relation between the X-ray QPO phenomenon
  and general relativity}, \emph{Nature}, \textbf{327}, p. 303.

\bibitem[{Paczy\'nski and Wiita(1980)}]{PW1980}
Paczy\'nski, B. and Wiita, P.~J. (1980), Thick accretion disks and superluminal
  luminosities, \emph{A\&A}, \textbf{88}, p.~23.

\bibitem[{van~der Klis~M.(2000)}]{vdk00}
van~der Klis~M. (2000), {Millisecond Oscillations in X-ray Binaries},
  \emph{AnnRevA\&A,}, \textbf{38}, p. 717.

\bibitem[{Woudt and Warner(2002)}]{woudt02}
Woudt, P.~A. and Warner, B. (2002), {Dwarf nova oscillations and quasi-periodic
  oscillations in cataclysmic variables - I. Observations of VW Hyi},
  \emph{MNRAS}, \textbf{333}, p. 411.

\end{thebibliography}

\end{document}